\documentclass[twocolumn]{aastex631}

\newcommand{\Jband}{$J-$ band}

\newcommand{\Kband}{$K_s-$ band}

\usepackage{subcaption}

\begin{document}

\title{The JWST Weather Report from the Isolated Exoplanet Analog SIMP 0136+0933: Pressure-Dependent Variability Driven by Multiple Mechanisms}

\correspondingauthor{Allison M. McCarthy}
\email{alliemc@bu.edu}

\author[0000-0003-2015-5029]{Allison M. McCarthy}
\affiliation{Department of Astronomy, Boston University, 725 Commonwealth Ave., Boston, MA 02215, USA}
\author[0000-0003-0489-1528]{Johanna M. Vos}
\affiliation{School of Physics, Trinity College Dublin, The University of Dublin, Dublin 2, Ireland}
\affiliation{Department of Astrophysics, American Museum of Natural History, Central Park West at 79th Street, New York, NY 10034, USA}

\author[0000-0002-0638-8822]{Philip S. Muirhead}
\affiliation{Department of Astronomy, Boston University, 725 Commonwealth Ave., Boston, MA 02215, USA}

\author[0000-0003-4614-7035]{Beth A. Biller}
\affiliation{Institute for Astronomy, University of Edinburgh, Royal Observatory, Edinburgh EH9 3HJ,UK}
\affiliation{Centre for Exoplanet Science, University of Edinburgh, Edinburgh, UK}

\author[0000-0002-4404-0456]{Caroline V. Morley}
\affiliation{Department of Astronomy, University of Texas at Austin, Austin, TX 78712, USA}

\author[0000-0001-6251-0573]{Jacqueline Faherty}
\affiliation{Department of Astrophysics, American Museum of Natural History, New York, NY 10024, USA}
:

\author[0000-0003-4600-5627]{Ben Burningham}
\affiliation{Centre for Astrophysics Research, School of Physics, Astronomy and Mathematics, University of Hertfordshire, Hatfield AL10 9AB, UK}

\author[0000-0002-2682-0790]{Emily Calamari}
\affiliation{Graduate Center of the City University of New York, New York, NY 10016, USA}
\affiliation{Department of Astrophysics, American Museum of Natural History, New York, NY 10024, USA}

\author[0000-0001-6129-5699]{Nicolas B. Cowan}
\affiliation{Department of Earth \& Planetary Sciences, McGill University, 3450 rue University, Montréal, QC H3A 0E8, Canada}
\affiliation{Department of Physics, McGill University, Montréal, QC H3A 2T8, Canada}

\author[0000-0002-1821-0650]{Kelle L. Cruz}
\affiliation{Department of Physics and Astronomy, Hunter College, City University of New York, 365 Fifth Avenue, New York, NY 10016, USA}
\affiliation{Graduate Center of the City University of New York, New York, NY 10016, USA}
\affiliation{Department of Astrophysics, American Museum of Natural History, New York, NY 10024, USA}

\author[0000-0003-4636-6676]{Eileen Gonzales}
\affiliation{Department of Physics and Astronomy, San Francisco State University, 1600 Holloway Ave., San Francisco, CA 94132, USA}

\author[0000-0002-9521-9798]{Mary Anne Limbach}
\affiliation{Department of Astronomy, University of Michigan, Ann Arbor, MI 48109, USA}

\author[0000-0001-7047-0874]{Pengyu Liu}
\affiliation{Institute for Astronomy, University of Edinburgh, Royal Observatory, Edinburgh EH9 3HJ,UK}
\affiliation{Centre for Exoplanet Science, University of Edinburgh, Edinburgh, UK}
\affiliation{Leiden Observatory, Leiden University, PO Box 9513, 2300 RA Leiden, The Netherlands}

\author[0000-0002-9792-3121]{Evert Nasedkin}
\affiliation{School of Physics, Trinity College Dublin, The University of Dublin, Dublin 2, Ireland}

\author[0000-0002-2011-4924]{Genaro Su\'arez}
\affiliation{Department of Astrophysics, American Museum of Natural History, New York, NY 10024, USA}

\author[0000-0003-2278-6932]{Xianyu Tan}
\affiliation{Tsung-Dao Lee Institute \& School of Physics and Astronomy, Shanghai Jiao Tong University, Shanghai 201210, People's Republic of China}

\author[0000-0002-1699-2222]{Cian O'Toole}
\affiliation{School of Physics, Trinity College Dublin, The University of Dublin, Dublin 2, Ireland}

\author[0000-0001-6627-6067]{Channon Visscher}
\affiliation{Chemistry \& Planetary Sciences, Dordt University, Sioux Center IA 51250}
\affiliation{Center for Extrasolar Planetary Systems, Space Science Institute, Boulder, CO 80301}

\author[0000-0001-8818-1544]{Niall Whiteford}
\affiliation{Department of Astrophysics, American Museum of Natural History, New York, NY 10024, USA}

\author[0000-0003-2969-6040]{Yifan Zhou}
\affiliation{Department of Astronomy, University of Virginia, 530 McCormick Rd., Charlottesville, VA 22904, USA}

%% Mark off the abstract in the ``abstract'' environment. 
\begin{abstract}

Isolated planetary-mass objects share their mass range with planets but do not orbit a star.  They lack the necessary mass to support fusion in their cores and thermally radiate their heat from formation as they cool, primarily at infrared wavelengths.  Many isolated planetary-mass objects show variations in their infrared brightness consistent with non-uniform atmospheric features modulated by their rotation. SIMP~J013656.5+093347.3 is a rapidly rotating isolated planetary-mass object, and previous infrared monitoring suggests complex atmospheric features rotating in and out of view. The physical nature of these features is not well understood, with clouds, temperature variations, thermochemical instabilities, and infrared-emitting aurora all proposed as contributing mechanisms. Here we report JWST time-resolved low-resolution spectroscopy from 0.8 -- 11 \micron~ of SIMP J013656.5+093347.3 which supports the presence of three specific features in the atmosphere: clouds, hot spots, and changing carbon chemistry. We show that no single mechanism can explain the variations in the time-resolved spectra. When combined with previous studies of this object indicating patchy clouds and aurorae, these measurements reveal the rich complexity of the atmosphere of SIMP J013656.5+093347.3. Gas giant planets in the solar system, specifically Jupiter and Saturn, also have multiple cloud layers and high-altitude hot spots, suggesting these phenomena are also present in worlds both within and beyond our solar-system.

\end{abstract}

%% Keywords should appear after the \end{abstract} command. 
%% The AAS Journals now uses Unified Astronomy Thesaurus concepts:
%% https://astrothesaurus.org
%% You will be asked to selected these concepts during the submission process
%% but this old "keyword" functionality is maintained in case authors want
%% to include these concepts in their preprints.


%% From the front matter, we move on to the body of the paper.
%% Sections are demarcated by \section and \subsection, respectively.
%% Observe the use of the LaTeX \label
%% command after the \subsection to give a symbolic KEY to the
%% subsection for cross-referencing in a \ref command.
%% You can use LaTeX's \ref and \label commands to keep track of
%% cross-references to sections, equations, tables, and figures.
%% That way, if you change the order of any elements, LaTeX will
%% automatically renumber them.
%%
%% We recommend that authors also use the natbib \citep
%% and \citet commands to identify citations.  The citations are
%% tied to the reference list via symbolic KEYs. The KEY corresponds
%% to the KEY in the \bibitem in the reference list below. 

\section{Introduction}

\label{sec:intro}

{Since its first observations in 2022, JWST has been transforming our understanding of extrasolar atmospheres. JWST's superb resolution and sensitivity is allowing detailed spectroscopic study across wavelengths previously only accessible with photometry or low-resolution spectroscopy. As such, JWST is revealing new insights into condensation and isotopic abundances \citep{Kuhnle2024}, disequilbrium chemistry \citep{Beiler2024a}, and possible auroral heating \citep{Faherty2024} in the faintest substellar objects. As well as allowing access to very faint atmospheres, the spectacular sensitivity of JWST allows high-SNR broad coverage spectra to be obtained with integrations short enough to allow detailed timeseries analysis, including wavelength regions that are essentially inaccessible from the ground. }
{These findings are reshaping our understanding of the formation, evolution, and structure of extrasolar atmospheres. Isolated planetary-mass objects, with their similar temperature, gravity, and chemistry to exoplanets, serve as analogs for directly-imaged exoplanets \citep{Faherty2016}. However, without the light of a host star, these isolated objects are easier to observe, making them ideal laboratories for studying atmospheric properties.}

{Photometric monitoring in the optical, near-IR, and mid-IR has uncovered dramatic variability that is common among brown dwarfs and isolated PMOs \citep[e.g.][]{Radigan2014,Metchev2015,Vos2022}. Inhomogeneous cloud cover \citep{Radigan2014}, thermochemical instabilities \citep{Tremblin2016}, hot spots \citep{2014ApJ...785..158R}, and auroral activity \citep{Hallinan2015} have all been suggested as drivers of this variability. These variability patterns have been linked to dynamical processes, such as cloud radiative feedback \citep{TanShowman2021a} and convective perturbations \citep{ZhangShowman2014}.} 

{Multiwavelength observations are crucial for probing different atmospheric layers and providing a comprehensive view of vertical structure \citep{Buenzli2012,Apai2013}. With JWST, we now have the capability for spectroscopic monitoring across an unprecedented wavelength range, essential for investigating atmospheric structures and processes in brown dwarfs, isolated planetary-mass objects, and directly imaged exoplanets. JWST's NIRSpec covers 0.6 to 5 ~\micron ~\citep{Jakobsen2022}, ideal for observing the brightest emission from these objects. JWST's MIRI extends this to 28 ~\micron, allowing exploration of the variability of mid-IR silicate absorption features for the first time \citep{Suarez2022}.}

{Recently, \citet{Biller2024} presented the first JWST spectroscopic monitoring of the brown dwarf binary WISE J104915.57-531906.1AB (WISE1049AB), detecting significant variability across 1–11 ~\micron. The wavelength-dependent light curve shapes were attributed to different pressure layers in the atmosphere, showcasing JWST’s ability to probe atmospheric structure in detail.}

{Here, we present the first JWST spectroscopic monitoring campaign of an isolated planetary-mass object: SIMP J013656.5+093347.3 (SIMP 0136+0933). As the brightest isolated planetary-mass object, it has been studied extensively with ground-based and space-based programs \citep[e.g.][]{Artigau2006, Artigau2009, Apai2013, Yang2016, McCarthy2024}. With an effective temperature of $\sim$1100 K (T2.5), it lies in the L/T transition region, known for high-amplitude variability driven by silicate cloud breakup \citep{Radigan2014, Liu2024}. SIMP 0136+0933 has shown variability amplitudes up to 5\% in the J band \citep{Artigau2009} and a rotation period of $\sim$2.4 h \citep{Croll, Yang2016}. Its near-IR variability has been linked to patchy clouds \citep{Apai2013, McCarthy2024}, while mid-IR variability may be driven by CO/CH$_4$ fingering convection \citep{Tremblin2020}. Strong pulsed radio emission on SIMP 0136+0933 indicates auroral activity \citep{Kao2016, Kao2018}.}

{In this paper, we present JWST observations of SIMP 0136+0933 (Section \ref{sec:observations}). We describe our reduced spectra and light curves in Section \ref{sec:results}. Section \ref{sec:Analysis} outlines our analysis, including light curve clustering and model interpretation. Finally, in Section \ref{sec:Conclusions}, we summarize our findings and suggest directions for future work on brown dwarfs and planetary-mass objects.}

\section{Observations and Data Reduction} \label{sec:observations}

As part of the JWST Cycle 2 program GO 3548 (PI: Vos), we observed just over one full rotation period of SIMP 0136+0933 sequentially with NIRSpec followed by MIRI. The NIRSpec and MIRI observations were carried out in an non-interruptible sequence since SIMP 0136+0933 is known to evolve significantly over the course of a few rotations \citep{Artigau2009}. This observing strategy ensured that we captured SIMP 0136+0933 in a similar atmospheric state across both instruments. NIRSPec Bright Object Time Series (BOTS) observations were carried out from UT 18:40:56 to 22:05:06 on Jul 23, 2023. MIRI Low Resolution Spectroscopy (LRS) Time Series Observations (TSOs) were then carried out from UT 22:05:11 on Jul 23, 2023 to UT 01:38:38 on Jul 24, 2023, with a brief background observation from UT 01:38:42 to UT 02:03:16 on Jul 24, 2023. 

We used the Prism/CLEAR mode for our NIRSpec/BOTS observations to obtain low-resolution time-series spectroscopy from $0.6-5.3~\mu$m with a resolution of $R=30-300$. WATA target acquisition was carried out with the F110W filter and the NRSRAPID readout pattern with 3 groups. For science observations, we used the SUB512 subarray and 7 groups per integration, yielding a cadence of 1.8~s to avoid saturation, obtaining a total of 5726 integrations. 

MIRI TSOs were obtained immediately after the NIRSpec/BOTS observations. We used the LRS slitless mode, the P750L disperser and FAST readout mode to provide maximal samples up the ramp, yielding time-resolved spectra at $5-14~\mu$m with a resolution of $R=40-160$. We used the F560W filter and the FASTR1 readout patterns with 5 groups per integration for target acquisition. For the science observations we then obtained a total of 575 integrations using the LRS SLITLESSPRISM subarray, FASTR1 readout and 120 groups per integration, yielding a cadence of 19.2~s. 

\subsection{NIRSpec Data Reduction}
The NIRSpec/BOTS data was reduced using the JWST STScI pipeline version 1.14.0, Calibration Reference Data System (CRDS) Version 11.17.19 and CRDS context file \texttt{jwst\_1253.pmap}.  The JWST Pipeline Stage 1 was used with default settings to apply basic detector-level corrections to all exposures. Following advice from the JWST Helpdesk, 1/f noise originating from JWST detector readout electronics was removed before running Stage 2. 
JWST NIRSpec readout electronics generate significant 1/f noise during detector operations and signal digitization. This noise varies from column to column, and appears as vertical banding that spans the entire width of the 2D spectral images in NIRSpec/BOTS observations, and can introduce systematic errors and significant scatter in light curves. We removed the 1/f noise by subtracting the median background flux per column across the spectral image. JWST Pipeline Stage 2 is then used to extract the 1D spectrum using an extraction width of 3 pixels. We run Stage 2 twice to produce the two final data products, one that is optimized for absolute flux and the other for relative light curves, for further analysis. To produce the final calibrated spectra for each image we run the Stage 2 pipeline using default parameters, which includes all steps. The final extracted near-infrared spectra are of excellent quality, with a median signal-to-noise ratio (SNR) of 64 per 1.8 s exposure. To produce relative light curves for variability analysis, we skip the flat field and photometric calibration steps, {following advice from the JWST helpdesk and recent examples of high-precision exoplanet transit and eclipse light curves from the literature \citep{Rustamkulov2022, Ahrer2023, Alderson2023, Sing2024, Welbanks2024}. Additionally, the curvature of the spectral trace is accounted for and 1/f noise is corrected in the relative light curve reduction, although this does not cause a significant difference in the light curves.}

\subsection{MIRI Data Reduction}
The MIRI/LRS time series observations were reduced using the same pipeline version and context file as the NIRSpec/BOTS data. Stage 1 processing was applied via \texttt{calwebb\_detector1} using default settings. The Stage 2 pipeline assigns wcs coordinates, source type, flat-fielding, photometric calibration and spectral extraction. The Stage 2 pipeline for TSOs does not include a background subtraction so we implement a custom background subtraction by defining a rectangular background region in the target image with the same width as the source extraction aperture and calculating the median value as a function of wavelength. The final extracted mid-infrared spectra are of good quality, with a median SNR of 124 per 19.2~s exposure.

\section{Results} \label{sec:results}

\subsection{Spectra} \label{sec:spectra}
\begin{figure*}
    \centering
    \includegraphics[width=0.9\textwidth]{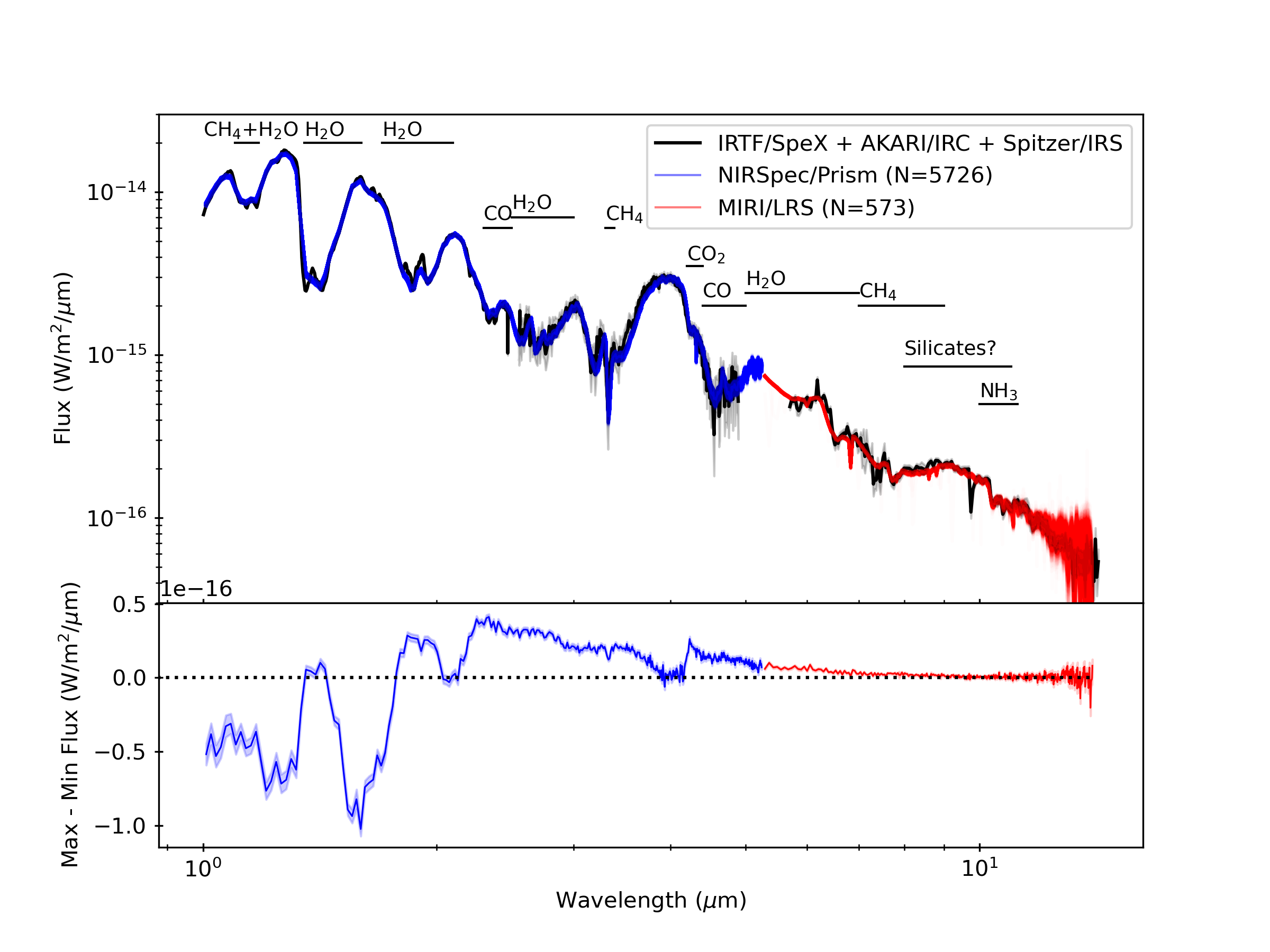}
    \caption{{Top panel: Final flux-calibrated NIRSpec/Prism (blue) and MIRI/LRS (red) spectra of SIMP 0136+0933, compared to the spectrum presented in \citet{Vos2023} (black) and error (shaded in grey). For the NIRSpec and MIRI spectra, each spectrum is plotted individually. Notable absorption features are indicated by horizontal black lines and their labels. For each instrument, we plot each spectrum individually, so the observed spread is a result of both noise and intrinsic variability.} {Bottom panel: Minimum spectrum subtracted from the maximum spectrum. The maximum and minimum spectra were identified using the $4.5-5.1~\mu$m light curve (Figure \ref{fig:overlap}).} }
    \label{fig:spectra}
\end{figure*}

In Figure \ref{fig:spectra} we present the final, flux-calibrated NIRSpec/PRISM and MIRI/LRS spectra for each exposure for SIMP 0136+0933. For both instruments, we plot each spectrum individually, so the observed spread shown in the top panel of Figure \ref{fig:spectra} is a result of both noise and intrinsic variability. {The bottom panel shows the difference between the ``maximum" and ``minimum" spectra, which is determined from the relative maximum and minimum of the 4.5 - 5.1~\micron~light curve shown in Figure \ref{fig:overlap}.} 

The JWST spectra are remarkably consistent with the spectrum analyzed by \citet{Vos2023}, which consists of three spectra (Infrared Telescope Facility (IRTF)/SpeX Prism, AKARI/Infrared Camera, and Spitzer/Infrared Spectrograph (IRS)) taken at different epochs. As expected for an early-T dwarf, we see absorption features driven by molecules such as H$_2$O, CH$_4$ and CO. The silicate absorption feature at $8.0-11~\mu$m is weak if present, which is consistent with its spectral type \citep{Suarez2022}. The spectrum within the silicate region aligns with the Spitzer/IRS spectrum analyzed by \citet{Vos2023}. 
Beyond wavelengths of $11~\mu$m, there is a noticeable drop in SNR which causes the increased spread beyond these wavelengths.

{The bottom panel of Figure \ref{fig:spectra} highlights that every wavelength is variable.
Additionally, the difference in the spectra demonstrates how each wavelength has unique temporal variations. If all wavelengths displayed the same overall variations, the difference between the ``maximum" and ``minimum" spectra would always be positive. Note that the ``maximum" and ``minimum" are selected from the 4.5 - 5.1 \micron light curves shown in Figure \ref{fig:overlap}. There is no global maximum or minimum, since the variability is extremely wavelength dependent.} A detailed spectral analysis, which will empirically measure molecular abundances and characterize the vertical cloud structure, will be presented in a future paper. This paper will focus on analysis of the relative light curves.

\subsection{Light Curves}

\begin{figure*}
    \centering
    \includegraphics[width=0.9\linewidth]{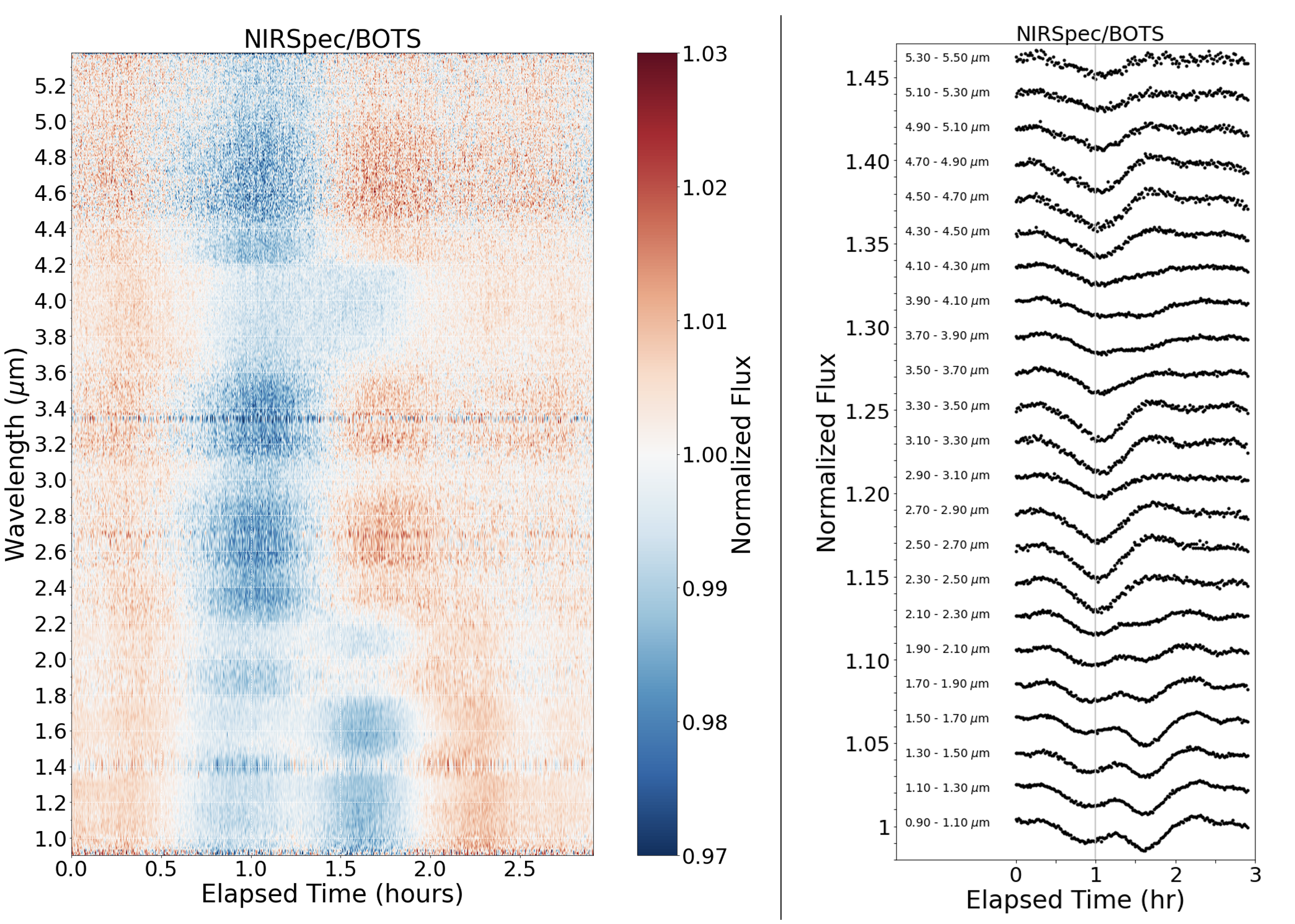}
    \caption{{{Left:} Variability map showing the reduced NIRSpec/BOTS spectrosopic relative light curves. The variability map is a 2-dimensional representation of the relative flux as a function of time and wavelength. {Right:} Binned light curves for NIRSpec. The data is binned by 0.2~\micron~ in wavelength and 1 minute in time. The gray vertical lines at 1 h marks the approximate minimum of the light curves, and denotes the ``start" of one full rotation (2.4 h) of the object.}}
    \label{fig:NVarandLC}
\end{figure*}

\begin{figure*}
    \centering
    \includegraphics[width=0.9\linewidth]{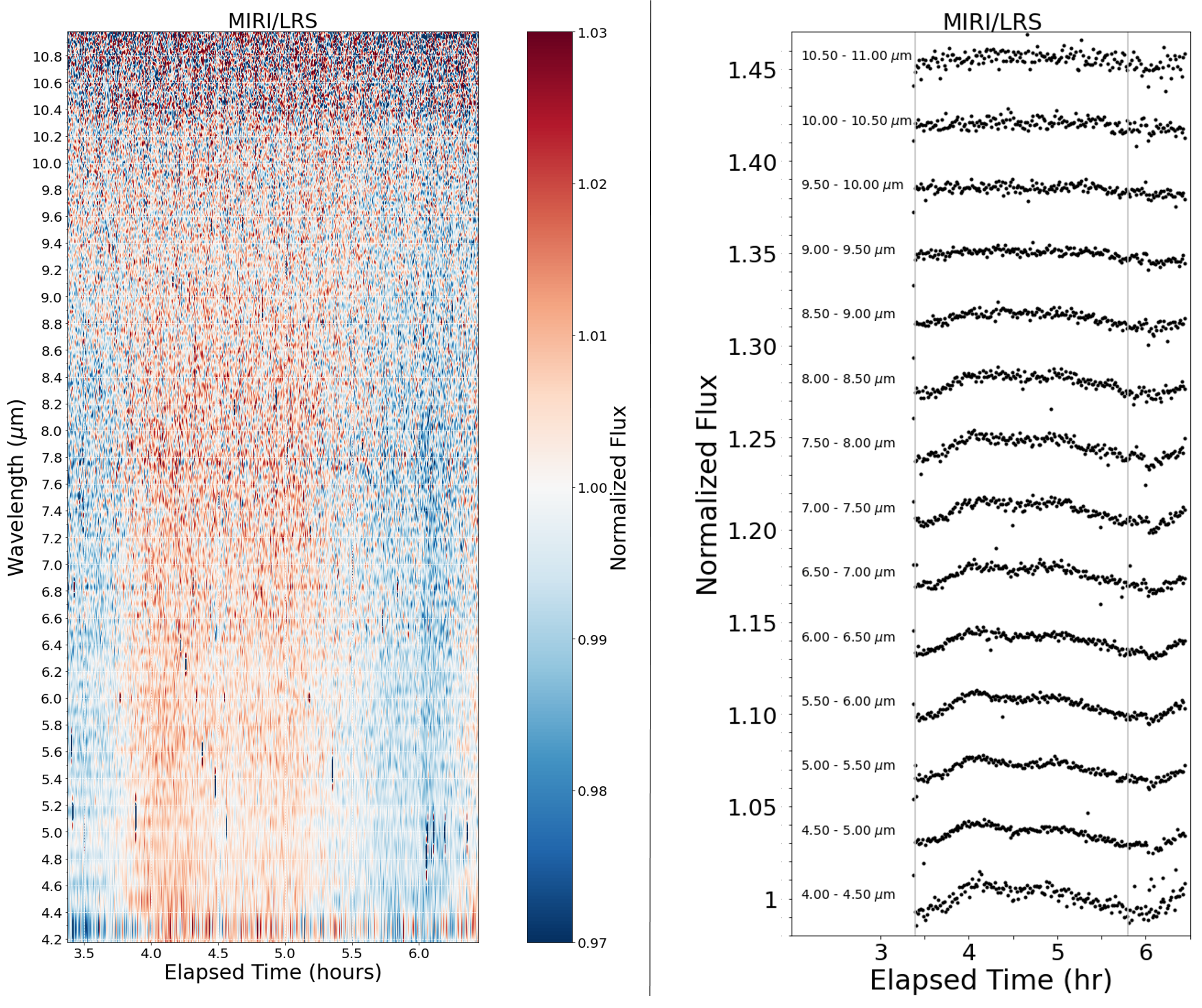}
    \caption{{{Left:} Variability map showing the reduced MIRI/LRS spectrosopic light curves. The variability map is a 2-dimensional representation of the relative flux as a function of time and wavelength. {Right:} Binned light curves for MIRI. The data is binned by 0.5~\micron~ in wavelength, and 1 minute in time. The gray vertical lines at 3.4 h, and 5.8 h, represent one complete rotation (2.4 h) of the object.}}
    \label{fig:MVarandLC}
\end{figure*}

We present variability maps of the NIRSpec and MIRI observations in Figures \ref{fig:NVarandLC} and \ref{fig:MVarandLC}. The variability map is a 2D representation of the normalized flux as a function of wavelength and time \citep[refer to ][for more details]{Biller2024}. The NIRSpec variability map interpolates over wavelengths which had poor SNR or clear anomalous points. The MIRI map includes all wavelengths, and the drop in SNR past $\sim$11~\micron~due to throughput, is clearly visible. Additionally, both the NIRSpec and MIRI wavelength resolution increases towards redder wavelengths. 

Figures \ref{fig:NVarandLC} and  \ref{fig:MVarandLC} display binned light curves for both the NIRSpec and MIRI data. NIRSpec light curves are binned in 0.2~\micron~ wavelength bins, and by 1 minute in time. MIRI light curves are binned by 0.5~\micron~ in wavelength, and by 1 minute in time. Prior to creating the binned light curves, we removed wavelengths whose light curves had an SNR$<25$ or displayed anomalous behaviour. 

Figures \ref{fig:NVarandLC} and  \ref{fig:MVarandLC} highlight the complex nature of the observed light curves as a function of wavelength, pointing to a dynamic atmosphere. The NIRSpec light curves shown in Figure \ref{fig:NVarandLC} are variable at all wavelengths and exhibit several distinct features. {At bluer wavelengths, there is a distinct double trough feature, where the relative depths of the two troughs varies as a function of wavelength. Moving towards the redder wavelengths, the light curve shape exhibits a single trough. From $\sim$3.5 -  4.3~$\mu$m the light curve shape changes again, before returning to one deep trough at the reddest of wavelengths.} 

\begin{figure}
    \centering
    \includegraphics[width=0.5\textwidth]{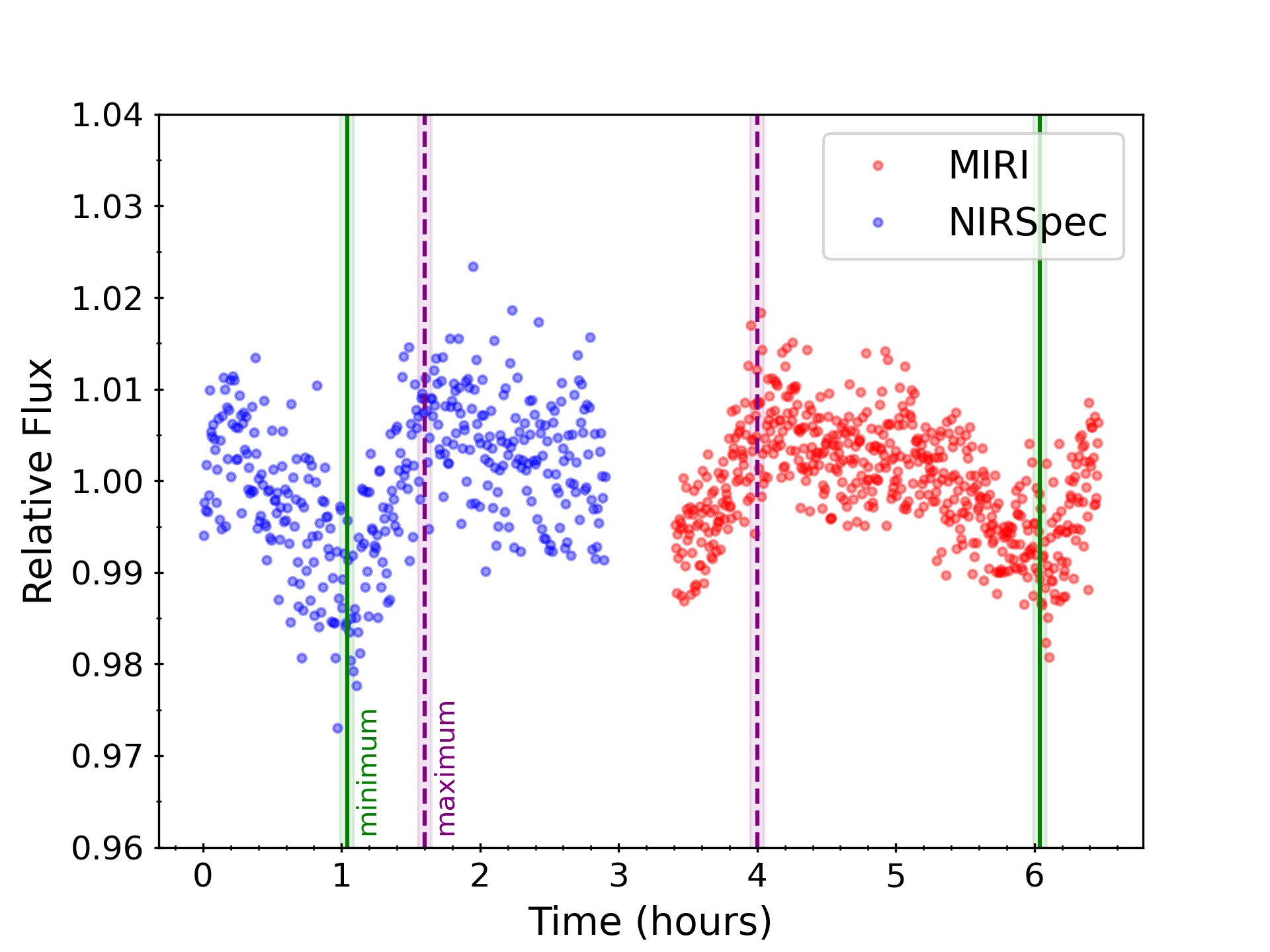}
    \caption{{NIRSpec and MIRI light curves in the overlapping wavelength regions ($4.5-5.1~\mu m$). The binned NIRSpec light curve ($27~$s cadence) is shown in blue and the MIRI light curve ($19~$s cadence) is shown in red. Green solid lines and purple dashed lines indicate timestamps for the ``maximum'' and ``minimum'' spectra referenced in Figure \ref{fig:spectra}.  }  }
    \label{fig:overlap}
\end{figure}

For the MIRI data, there are also patterns in the data. Overall, MIRI displays a maximum deviation of variability, the difference between the maximum and minimum normalized flux values, that is less than the variability observed in NIRSpec. From $\sim$4.2 -- $\sim$8.5 \micron, there is a double peaked feature, with the first peak at $\sim$4.0 h, and the secondary peak at $\sim$ 5.0 h (also shown in Figure \ref{fig:overlap}). At $\gtrsim$ 8.5 ~\micron, the behavior changes to a smoother light curve behavior with smaller maximum deviation. 

NIRSpec and MIRI share a region of wavelength space that allows us to create a light curve that covers the full observation duration. We plot light curves spanning this $4.5-5.1~\mu $m region in Figure \ref{fig:overlap}. We identify a minimum and maximum within the NIRSpec and MIRI portion of the light curve by eye, which is shown by the solid green and dashed purple vertical lines in Figure \ref{fig:overlap}.  We find that, at these wavelengths, the light curve does not change significantly from one rotation to the next \citep[Compare to][who found evolution in light curve structure from rotation to rotation]{Biller2024}. Spectroscopic observations spanning longer than 2 rotations will be necessary to probe in detail the long-term evolution of the atmosphere of SIMP 0136+0933.

\section{Analysis and Discussion} \label{sec:Analysis}
\subsection{Light Curve Modeling} \label{sec:CeleriteModeling}

We used {\tt celerite2} \citep{celerite1, celerite2} to fit a model to each of our NIRSpec and MIRI light curves, following the approach of \citet{McCarthy2024}. The {\tt celerite2} software employs Gaussian processes to model data as correlated noise, with the degree of correlation set by a chosen kernel function \citep{Rasmussen2006Gaussian}.  We used a combination of two \texttt{SHOTerm} kernels which represent stochastically driven, damped simple harmonic oscillators with a fixed period matched to the measured rotational period of SIMP 0136+0933 \citep[2.4 h][]{Artigau2009,Yang2016}.  Together, the kernels have six free parameters, and we included one additional parameter to capture additional measurement uncertainty following a recommended approach in the {\tt celerite2} documentation.  We then used the \texttt{emcee} Markov Chain Monte Carlo software package to explore the values of the free parameters that best describe the data \citep{emcee}.  The set of free parameters with the highest posterior value was used to calculate the best-fitting model.  When fitting the light curves, we did not bin the NIRSpec data in wavelength or time. We binned the MIRI light curves by 0.5~\micron{} to increase the signal-to-noise. The {\tt celerite2} fits are shown in gray in panel (a) of  Figures \ref{fig:NIRSpecBigPlot} and \ref{fig:MIRIBigPlot}.

\subsection{Light Curve Clustering}\label{sec:cluster}

The variability maps and light curves presented in Figures \ref{fig:NVarandLC} and \ref{fig:MVarandLC}, show complex light curve behavior. We observe variability across all wavelengths (also discussed in Section \ref{sec:Amp}), with similar behaviors seen across distinct wavelength regions. {Wavelengths with similar light curves are likely impacted by the same mechanism(s), so by grouping together similar shaped light curves we can investigate these mechanism(s).}

{We use a $K$-means clustering algorithm from \texttt{scikit-learn} to efficiently group together light curves with similar features, capturing details difficult to discern visually, using the approach employed by \citet{Biller2024}.
We bin every 50 points from the {\tt celerite2} highest likelihood fits, which results in a cadence of $\sim$55s for NIRSpec, which we feed to the $K$-means clustering algorithm.} We do not bin the NIRSpec data in wavelength. For the MIRI data, since we have already binned the data in wavelength space, we do not bin it in time for the $K$-means clustering algorithm.

After making cuts where the average SNR $<25$ and at wavelengths where there were clear anomalous points in the data, we have 362 wavelengths in NIRSpec, producing the same number of light curves. After binning the spectra by 0.5 \micron, and making the same SNR cuts as NIRSpec, the MIRI data results in 14 light curves. In each case, we allow the $K$-means clustering algorithm to determine the optimal number of clusters using the \texttt{KneeLocator} from \texttt{kneed} \citep{KneeLocator}. \texttt{KneeLocator} applies the elbow-method, which uses the sum-of-square-errors to identify the point at which an additional cluster provides diminishing return. For NIRSpec, we allow the clustering algorithm to select between 1 and 80 clusters, and for MIRI between 1 and 10 \citep[Compare to][who allowed $n_{cluster}$ to vary between 1 and 11 for both NIRSpec and MIRI]{Biller2024}. This difference in number of allowable clusters between NIRSpec and MIRI is due to the binning of the MIRI data which resulted in only 14 light curves. The $K$-means clustering algorithm determines that the optimal number of clusters is 9 for NIRSpec and 2 for MIRI. This is an intriguing difference from \citet{Biller2024} who found 3 clusters for NIRSpec and 2 for MIRI for both components of the binary, despite both components having a similar effective temperature to SIMP 0136+0933. 

At a first glance, we recognize that NIRSpec clusters 1 -- 5, 6 and 7, and 8 and 9 appear similar (Figure \ref{fig:NIRSpecBigPlot}). {In addition to the clustering algorithm identifying these clusters as distinct, we validated their uniqueness by conducting cosine similarity and Pearson correlation coefficient tests.} We discuss the more nuanced differences between clusters, and what may cause them in, Sections \ref{sec:pressurecluster} and \ref{sec:Amp}.

\subsection{Pressure Probed by Light Curve Clusters can Inform the Primary Mechanism of Variability}\label{sec:pressurecluster}

\begin{figure*}
    \centering
    \includegraphics[width=0.99\textwidth]{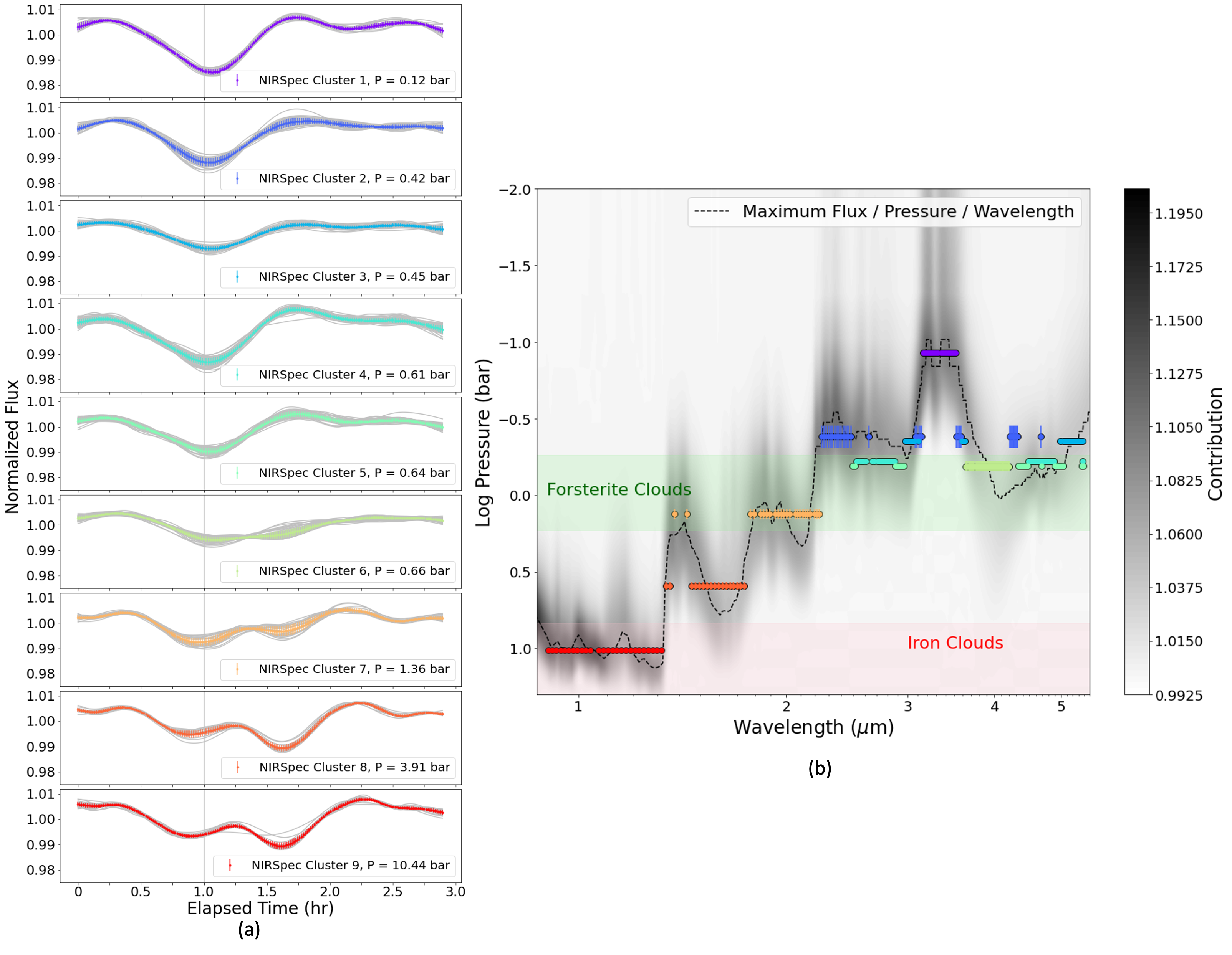}
    \caption{Panel (a) shows the 9 clusters obtained from the NIRSpec light curves. The celerite fit for each light curve that comprises each cluster is shown in gray, and the average for each cluster is shown in the colored marker. The gray vertical lines in panel (a) of both Figures \ref{fig:NIRSpecBigPlot} and \ref{fig:MIRIBigPlot} mark the 2.4 h rotation period of this object. Panel (b) shows the Sonora Diamondback contribution function for a clear atmosphere with $T_{\rm eff}=1100$K and log(g)=4.5. The colors are the same for the same clusters across both subplots. The clusters are overlaid onto the contribution function by calculating the average pressure of maximum flux for the wavelengths which comprise each cluster. The error bars on the clusters in panel (b) are the variance for the pressure of each cluster. The green shaded region represents the pressures where forsterite clouds exist, and the pink shaded region represents the pressures where iron clouds might exist from \citet{Vos2023}.}
    \label{fig:NIRSpecBigPlot}
\end{figure*}

\begin{figure*}
    \centering
    \includegraphics[width=0.99\textwidth]{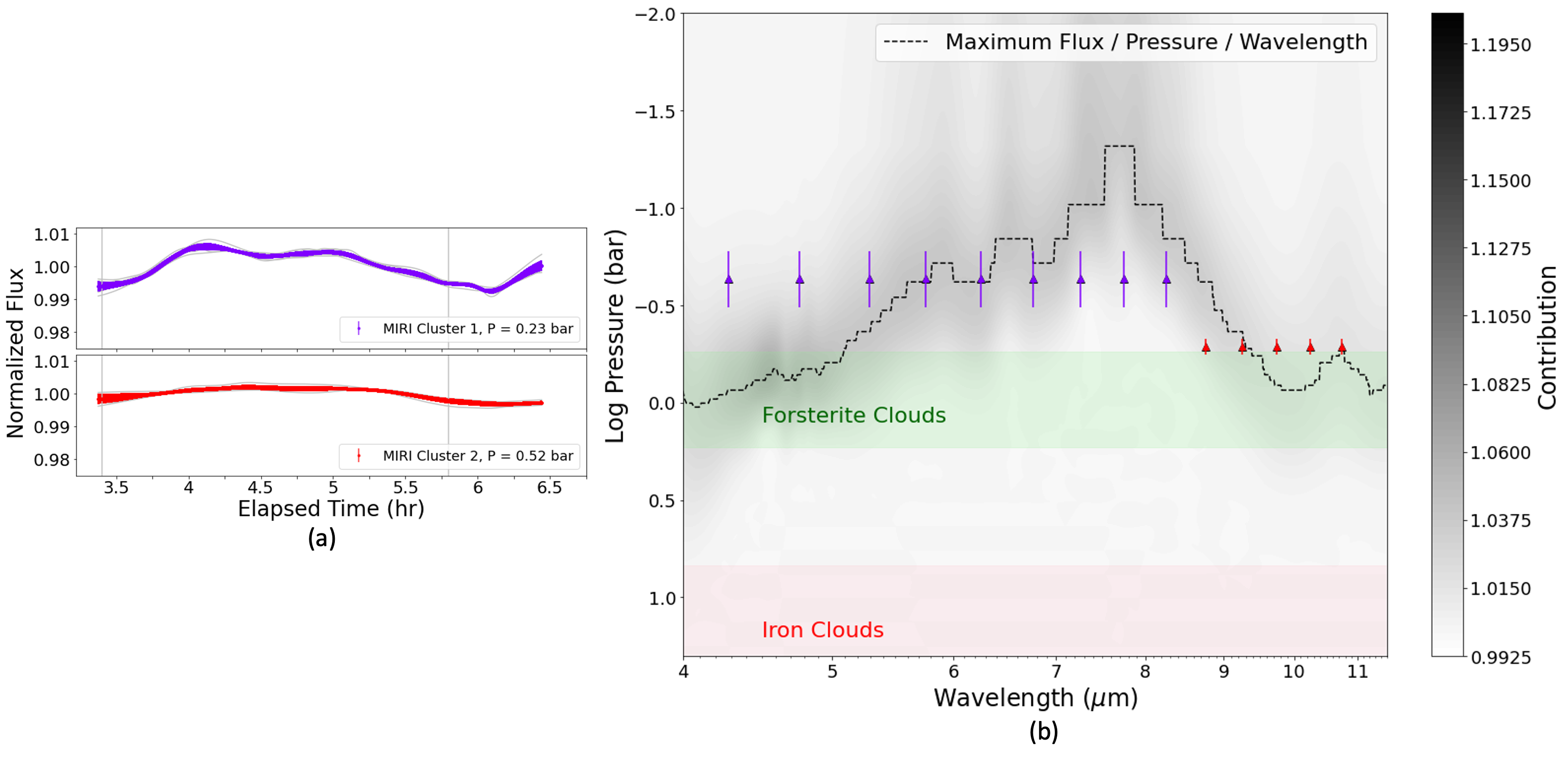}
    \caption{Panel (a) shows the 2 clusters obtained from the MIRI light curves. The celerite fit for each light curve that comprises each cluster is shown in gray, and the average for each cluster is shown in the colored marker. The gray vertical lines in panel (a) of both Figures \ref{fig:NIRSpecBigPlot} and \ref{fig:MIRIBigPlot} mark the 2.4 h rotation period of this object. Panel (b) shows the Sonora Diamondback contribution function for a clear atmosphere with $T_{\rm eff}=1100$K and log(g)=4.5. The colors are the same for the same clusters across both subplots. The clusters are overlaid onto the contribution function by calculating the average pressure of maximum flux for the wavelengths which comprise each cluster. The error bars on the clusters in panel (b) are the variance for the pressure of each cluster. The green shaded region represents the pressures where forsterite clouds exist, and the pink shaded region represents the pressures where iron clouds might exist from \citet{Vos2023}.}
    \label{fig:MIRIBigPlot}
\end{figure*}

{The clustering algorithm reveals which wavelengths have similarly shaped light curves. Since physical mechanisms like clouds, aurora, hot spots, or chemical instabilities may all cause light curve variability, we infer that similar shapes indicate shared physical mechanism(s). Figures \ref{fig:NIRSpecBigPlot} and \ref{fig:MIRIBigPlot} panel (a) show the clustered light curves, with gray vertical lines marking the object's 2.4 h rotation period. The line in Figure \ref{fig:NIRSpecBigPlot} panel (a) is at 1 h, and in Figure \ref{fig:MIRIBigPlot} panel (a) at 3.4 and 5.8 h.}

{We pair the multi-wavelength clustering information with atmospheric models to provide insights into the atmosphere's vertical structure. Using the Sonora Diamondback models \citep{Morley2024}, we match each wavelength to its pressure/temperature layer. Figures \ref{fig:NIRSpecBigPlot} and \ref{fig:MIRIBigPlot} panel (b) overlay the NIRSpec and MIRI clusters on a clear-atmosphere contribution function at $T_{\mathrm{eff}}=$ 1100,K and $\log(g)=4.5$ \citep[similar to SIMP 0136+0933,][]{Gagne2017,Vos2023}. We use a cloudless contribution function, because a clear atmosphere shows the deepest observable level in the atmosphere. The addition of inhomogeneous features would alter this function only to reveal shallower depths, not deeper. For instance, clouds flatten the contribution function at their formation pressure \citep[see Figure 5 in][]{Vos2023}.} We determine the average pressure probed for each cluster by determining the pressure which has the maximum flux contribution emitted by each wavelength in the cluster. This is shown in figures \ref{fig:NIRSpecBigPlot} and \ref{fig:MIRIBigPlot} with error bars showing the pressure variance.

{Prior works suggest physical mechanisms that may be responsible for SIMP 0136+0933’s variability. A retrieval analysis by \citet{Vos2023} finds forsterite clouds at 0.55–1.7 bar and an iron cloud layer that is optically thick at ~7 bar, shown as green and pink shaded regions in panel (b) of Figures \ref{fig:NIRSpecBigPlot} and \ref{fig:MIRIBigPlot}. The presence of an iron cloud at 7 bar would block flux from deeper pressures in the 0.88-1.32~\micron~ wavelength range, which is matched by NIRSpec Cluster 9 (red markers). Likewise, a forsterite cloud at 0.55–1.7 bar would block flux from that region or below, with wavelengths 0.81–2.17~\micron (NIRSpec clusters 7–9) reflecting variability from forsterite cloud inhomogeneity. The forsterite cloud spans 0.55–1.7 bar, with clouds becoming optically thick at different pressures per wavelength, impacting flux in NIRSpec cluster 6. MIRI cluster 2’s wavelengths probe the very top of the forsterite cloud layer and as such show low variability amplitudes compared to other clusters which probe a majority or the entirety of the forsterite cloud layer.}

{While clouds can explain variability in NIRSpec clusters 7–9, they alone cannot account for the variability observed in clusters 1–6. Clusters 1–3 probe higher than silicate cloud levels, while clusters 4–6 only probe the very top of the forsterite cloud. Since models show no clouds at pressures shallower than 0.55 bar, additional sources are needed for NIRSpec clusters 1–6. While \citet{Vos2023} did not consider a hot spot, \citet{Faherty2024} suggests that an aurorally-driven temperature inversion may drive methane emission in a Y dwarf atmosphere. Strong aurorae in SIMP 0136+0933 \citep{Kao2016, Kao2018} suggest that an aurorally-driven temperature inversion may be plausible in this case too. Beyond aurora, rapid dynamical mixing may give rise to hot spots \citep{Morley2014}, and deeper temperature perturbations may be transported to lower pressures via radiative heating \citep{RobinsonMarley2014, Tremblin2020}.}

{We compare to the model spectra of \citet{Morley2014} in Figure \ref{fig:ampwave}, which inject energy at 0.1 bar and could represent any heating mechanism. Further work is needed to link specific mechanisms, like auroral heating, to spectral changes. The \citet{Morley2014} spectra show that a hot spot impacts wavelengths 2.2–3.8 \micron and 5.4–8.5 \micron, corresponding to NIRSpec cluster 1 and portions of clusters 2–6, as well as MIRI cluster 1. Wavelengths with the highest flux ratios, 2.6 and 3.2 \micron, align with the maximum observed deviations, discussed in \ref{sec:Amp}.}

{Finally, our analysis shows that NIRSpec clusters 2–6 probe similar pressures (0.42–0.66 bar) despite different light curve shapes. We suspect these shape differences arise from the abundance of specific molecules. Labels for CO, CO${\rm _2}$, CH${\rm _4}$, and H${\rm _2}$O (Figure \ref{fig:ampwave} bottom panel, Section \ref{sec:Amp}) show which wavelengths are impacted by each molecule. For instance, NIRSpec cluster 1 traces the methane band from 3.2–3.7 \micron, as predicted by \citet{Tremblin2020}. Clusters 2–6 are more challenging to assign due to overlapping cross-sections.}

{It is likely that each cluster is impacted by a combination of mechanisms, but at varying ratios. For this reason, we cannot rule out degeneracies in the combination of multiple mechanisms. For example, two clusters may probe pressures that are impacted by both clouds and a hot spot, but in one cluster, the clouds are the primary mechanism, and in another cluster the hot spot is the primary mechanism, although the clouds still impact the light curve. The difference in primary mechanism is due to wavelengths affected by each mechanism (Figure \ref{fig:ampwave}).}

\subsection{Maximum Deviation as a Function of Wavelength}\label{sec:Amp}

\begin{figure}
    \centering
    \includegraphics[width=0.99\columnwidth]{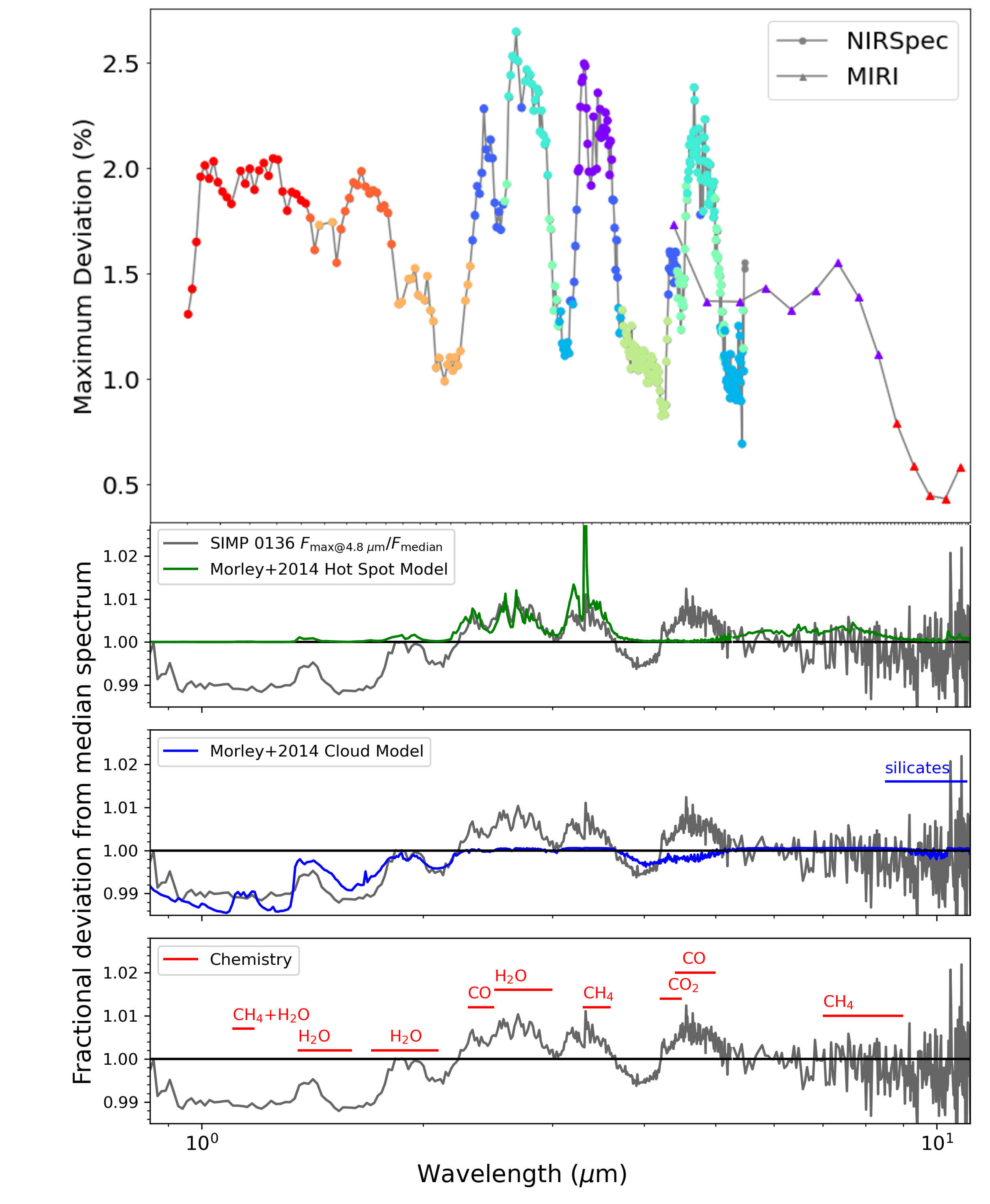}
    \caption{{The top panel shows the measured maximum deviation of the light curves, colored by their respective cluster. The NIRSpec data are shown as circle markers, and the MIRI data, triangles.} { The gray spectra in the three sub panels show the deviation of our ``maximum" spectrum from our median spectra. The second panel shows the comparison between a scaled model flux ratio between a clear atmosphere and a clear atmosphere with a portion covered by a hot spot \citep{Morley2014}. The third panel highlights the flux ratio between a cloudy and clear atmosphere \citep{Morley2014}.  The final panel labels the locations of impact for abundant molecules in the atmosphere.}}
    \label{fig:ampwave}
\end{figure}

{The top panel of Figure \ref{fig:ampwave} shows the maximum observed flux deviation as a function of wavelength. Given the non-sinusoidal nature of these light curves, we define maximum flux deviation as the difference between the maximum and minimum normalized flux values of the celerite maximum likelihood fits. 
Notably, every wavelength displays variability. The maximum deviations occur near $\sim$2.6 and $\sim$3.2 \micron. According to the contribution function in Figure \ref{fig:NIRSpecBigPlot}, these wavelengths probe some of the highest altitudes in SIMP 0136+0933 and align with those most affected by an upper-atmosphere hot spot, as shown in the second panel of Figure \ref{fig:ampwave}.} {We do not carry out a full spectral modelling analysis in this paper, but simply highlight the similarities between the observations and the predictions from \citep{Morley2014}. Future atmospheric retrievals will better answer these questions.} 

{Between $\sim$5.5 and $\sim$8.5 \micron, we observe larger amplitudes than at wavelengths beyond $\sim$8.5 \micron, consistent with predictions from the \citet{Morley2014} hot spot model. These amplitude patterns support the presence of a hot spot in the upper atmosphere.}

{The second subpanel of Figure \ref{fig:ampwave} shows flux ratio predictions driven by patchy clouds from \citet{Morley2014}.  $\sim$1 to $\sim$2.2 and $\sim$3.7 to $\sim$5 show a decreased flux due to the presence of clouds. These wavelengths roughly correspond to NIRSpec clusters 6-9, which all display a similar double trough shape.}

{A secondary amplitude peak appears at $\sim$4.6 \micron, not explained by the \citet{Morley2014} hot spot or cloud models. However, this peak aligns with the carbon dioxide bandpass (4.2–5.0 \micron) and the carbon monoxide bandpass (4.4–5.2 \micron), as shown in the bottom panel of Figure \ref{fig:ampwave}. Notably, NIRSpec clusters 1 - 5 (spanning $\sim$2.2–3.7 \micron and $\sim$4.2–5.2 \micron) all show a distinct trough near $\sim$1.05 h, suggesting a shared underlying mechanism for this feature. Since hot spot activity may drive variability from $\sim$2.2 to $\sim$3.7 \micron, we infer that it may influence variability from $\sim$4.2 to $\sim$5.2 \micron. Any disequilibrium species in the upper atmosphere would also vary in-phase with the hot spot because the added energy from the hot spot could facilitate chemical interactions. In-depth atmospheric modeling, including both a hot spot model and disequilibrium chemistry, will be necessary to assess the impact of the hot spot on carbon chemistry within the atmosphere.}

{At a glance, NIRSpec clusters 1–5 have similar shapes, due to a prominent trough near $\sim$1 h. However, detailed inspection shows significant differences in light curve shapes beyond 2 h, likely due to varying contributions from abundant molecules such as methane, water, carbon monoxide, and carbon dioxide \citep[c.f.][]{Fortney2020}.} {The impact of changing carbon chemistry is still an open question in the field, and further Global Circulation Modeling (GCM) efforts with disequilibrium chemistry will be needed to understand the scope of this variability mechanism \citep{Lee2023,Lee2024}.}

{Finally, MIRI cluster 2 (8.5–11 \micron), shows less than 1\% amplitude variability. While these wavelengths correspond to the silicate absorption region, the expected variability is low compared to the \Jband~\citep[Figure 7,][]{Vos2023}. However, \citet{Luna2021} predicts that wavelengths beyond 9 \micron could exhibit enhanced variability due to patchy, high-altitude silicate clouds with small grains. In our analysis, we attribute increased variability from 5 to 8.5 \micron to the hot spot presence. Similar reduced variability beyond 8.5 \micron is observed by \citet{Biller2024} for WISE1049AB, likely due to silicate absorption properties, which are prominent from L4–L6 but absent by L8 \citep{Suarez2022}. In T dwarfs, forsterite clouds form deeper in the atmosphere than in their warmer L dwarf counterparts. In this case, the wavelengths which make up MIRI cluster 2 probe the very top of the forsterite cloud layer, and as such any inhomogeneity in the cloud layer has minimal impact in the variations in brightness. }

\subsection{Comparison with previous studies}

{SIMP 0136 has been extensively studied using a range of observational and analysis techniques. \citet{Apai2013} obtained HST/WFC3  spectroscopic monitoring of SIMP 0136+0933, reporting a 4.5\% amplitude—higher than our maximum flux deviation across all wavelengths. They also observed significant light curve evolution. \citet{Yang2016} conducted simultaneous Spitzer and HST observations, finding wavelength-dependent differences in light curve shape and amplitude, consistent with our results. In their study, the \Jband~ light curve had an amplitude of $\sim$5\%, greater than our $\sim$2\% maximum deviation, while Spitzer showed $\sim$1\% amplitude, lower than our observed deviations \citep[Figure 7, ][]{Yang2016}. They reported a $\sim$30$^{\circ}$ phase shift between HST \Jband~ and Spitzer channel one but found no phase shift in and out of the water bands. They attributed different light curve shapes to wavelengths probing opposite sides of $\sim$4 bar, near the radiative-convective boundary. \citet{McCarthy2024} identified a $\sim$40$^{\circ}$ phase shift between \Jband~ and \Kband, linking it to multiple patchy cloud layers. Similarly, \citet{Plummer2024} reported a $\sim$90$^{\circ}$ phase shift between $H-K$ and $J-H$ light curves, suggesting complex vertical structure and wave-induced cloud breakup in SIMP 0136’s atmosphere.}

{Previous studies \citep[e.g.][]{Buenzli2012, Biller2013, Yang2016, Lew2020, McCarthy2024} have noted phase shifts between light curves in different photometric bands for many brown dwarfs and planetary-mass objects, including SIMP 0136. However, our results indicate that rather than a true phase shift, distinct light curve shapes arise from different variability mechanisms present at varying depths (pressure/temperature levels). Broadband photometric observations may capture multiple mechanisms, creating the illusion of a phase shift in the integrated light curve arising from the superposition of multiple variability mechanisms. Since JWST provides both extensive wavelength range with adequate resolution, we can discern these distinct light curve shapes, allowing us to uncover the drivers of variability.}

\section{Summary and conclusions} \label{sec:Conclusions}

{This study, the first JWST spectroscopic variability analysis of a planetary-mass object, demonstrates JWST’s unique power to probe extrasolar atmospheres. As a young exoplanet analog, SIMP 0136+0933 is an ideal laboratory for studying the diversity of atmospheric variability mechanisms from 1–12 \micron. The results highlight the potential of combining advanced atmospheric models with JWST spectroscopic observations and motivate further JWST variability studies of extrasolar atmospheres.}

{We report spectroscopic monitoring results from $\sim$3 h of NIRSpec/BOTS followed by $\sim$3 h of MIRI/LRS of the planetary-mass object SIMP 0136+0933 (rotation period 2.4 h). Variability is seen at all wavelengths, with the largest maximum flux deviation of 2.6\% at 2.6 \micron. The light curve structure changes across wavelength and pressure, indicating multiple atmospheric variability mechanisms.}

{Atmospheric features like clouds, aurora, hot spots, a reduced temperature gradient, or changing carbon chemistry could impact flux at specific wavelengths. Grouping light curves by shape, we identify which wavelengths are impacted by the same mechanism(s). We find 9 clusters in NIRSpec and 2 in MIRI, linked to patchy clouds, hot spots, and varying carbon chemistry.}

{Previous studies \citep{Apai2013,Vos2023} suggest cloud layers at $\sim$1 bar and $\sim$7 bar, impacting wavelengths from 0.88–2.17 \micron, corresponding to NIRSpec clusters 7, 8, and 9. Wavelengths beyond $\sim$2.2 \micron~correspond to NIRSpec clusters 1–6, showing broadly similar light curve shapes but different substructure beyond 2 h. A hot spot, possibly from aurora \citep{Kao2016,Kao2018}, rising/falling air pockets \citep{Morley2014}, or deeper temperature perturbations \citep{RobinsonMarley2014}, likely affects wavelengths $\sim$2.2–3.7 \micron ~\citep[Figure 4 in][]{Morley2014}. These correspond to NIRSpec cluster 1 and parts of clusters 2–6. Furthermore, changes in methane, water, carbon monoxide, and carbon dioxide likely cause variability which then causes impacted wavelengths' light curves to differentiate into distinct clusters.}

{ The two MIRI clusters are split at $\sim$8.5 \micron, coinciding with the end of hot spot-driven variations \citep[$\sim$5.5–8.5 \micron, Figure 4 in][]{Morley2014}, and the onset of the silicate absorption feature \citep{Suarez2022}, suggesting sensitivity to these mechanisms.}

{However, this is just a glimpse into SIMP 0136+0933's atmosphere. Like planets in our solar system, its atmosphere exhibits long-term changes \citep{Yang2016,Croll}. Longer observations spanning multiple rotations are crucial for deeper insights into evolving atmospheric mechanisms and distinguishing variability timescales. Monitoring changes in light curve shapes over time will help disentangle these mechanisms and assess their correlations.}

{This is the first paper in a series exploring the data from Section \ref{sec:observations}. Detailed modeling, retrievals, and spectral analysis will follow, offering a deeper understanding of the physical drivers of variability on this extrasolar world.}

All JWST data used in this paper can be found in MAST: \dataset[10.17909/pfnd-md36]{http://dx.doi.org/10.17909/pfnd-md36}.

\begin{acknowledgments}

We thank the referee for their helpful comments, which improved the quality of this publication. This work is based on observations made with the NASA/ESA/CSA James Webb Space Telescope. The data were obtained from the Mikulski Archive for Space Telescopes at the Space Telescope Science Institute, which is operated by the Association of Universities for Research in Astronomy, Inc., under NASA contract NAS 5-03127 for JWST. These observations are associated with program GO:3548. This work was supported in part by JWST-GO-03548.004-A.

A. M. M. acknowledges support from the National Science Foundation Graduate Research Fellowship Program under Grant No. DGE-1840990. 

J. M. V. would like to thank Jennifer Kestell and Mario Fabelo Ozcariz for helpful discussions regarding the JWST data. J. M. V., C. O'T. and E. N. acknowledge support from a Royal Society - Science Foundation Ireland University Research Fellowship (URF$\backslash$R1$\backslash$221932, URF$\backslash$ERE$\backslash$221108). 
BB acknowledges support from UK Research and Innovation Science and Technology Facilities Council [ST/X001091/1].
C.V. acknowledges support from JWST cycle 1 GO AR theory program PID-2232.

N.B.C.\ acknowledges support from an NSERC Discovery Grant, a Tier 2 Canada Research Chair, and an Arthur B.\ McDonald Fellowship, and thanks the Trottier Space Institute and l’Institut de recherche sur les exoplanètes for their financial support and dynamic intellectual environment.

\end{acknowledgments}

%% To help institutions obtain information on the effectiveness of their 
%% telescopes the AAS Journals has created a group of keywords for telescope 
%% facilities.
%
%% Following the acknowledgments section, use the following syntax and the
%% \facility{} or \facilities{} macros to list the keywords of facilities used 
%% in the research for the paper.  Each keyword is check against the master 
%% list during copy editing.  Individual instruments can be provided in 
%% parentheses, after the keyword, but they are not verified.

\vspace{5mm}
\facilities{JWST (NIRSpec, MIRI)}

%% Similar to \facility{}, there is the optional \software command to allow 
%% authors a place to specify which programs were used during the creation of 
%% the manuscript. Authors should list each code and include either a
%% citation or url to the code inside ()s when available.

\software{{\tt astropy} \citep{astropy:2022},  
          {\tt SciPy} \citep{SciPy-NMeth2020}, 
          {\tt celerite2} \citep{celerite2},
          {\tt emcee} \citep{Foreman-Mackey2013}
}

%% Appendix material should be preceded with a single \appendix command.
%% There should be a \section command for each appendix. Mark appendix
%% subsections with the same markup you use in the main body of the paper.

%% Each Appendix (indicated with \section) will be lettered A, B, C, etc.
%% The equation counter will reset when it encounters the \appendix
%% command and will number appendix equations (A1), (A2), etc. The
%% Figure and Table counter will not reset.

%% For this sample we use BibTeX plus aasjournals.bst to generate the
%% the bibliography. The sample631.bib file was populated from ADS. To
%% get the citations to show in the compiled file do the following:
%%
%% pdflatex sample631.tex
%% bibtext sample631
%% pdflatex sample631.tex
%% pdflatex sample631.tex

\bibliography{references}{}
\bibliographystyle{aasjournal}

%% This command is needed to show the entire author+affiliation list when
%% the collaboration and author truncation commands are used.  It has to
%% go at the end of the manuscript.
%\allauthors

%% Include this line if you are using the \added, \replaced, \deleted
%% commands to see a summary list of all changes at the end of the article.
%\listofchanges

\end{document}